\documentstyle{article}
\newcommand{\be}{\begin{equation}}
\newcommand{\ee}{\end{equation}}

\begin{document}


\title
{\Large {\bf 
Study of a Class of Four Dimensional Nonsingular 
Cosmological Bounces}}
\author{Fabio Finelli \thanks{Also supported by INFN, Sezione di 
Bologna, via Irnerio 46 -- I-40126 Bologna -- Italy} \\
IASF/CNR, Istituto di Astrofisica Spaziale e
Fisica Cosmica \\ Sezione di Bologna \\ Consiglio Nazionale delle 
Ricerche \\
via Gobetti, 101 -- I-40129
Bologna -- Italy}
\date{\today}
\maketitle
\begin{abstract}
We study a novel class of nonsingular time-symmetric cosmological bounces. 
In this class of four dimensional models the bounce is induced by a 
perfect fluid with a negative energy density. 
Metric perturbations are solved in an analytic way all through the bounce. 
The conditions for generating a scale invariant spectrum of tensor 
and scalar metric perturbations are discussed.
\end{abstract}


\vskip 0.4cm

\section{Introduction}

Bouncing universes become attractive and fashionable periodically. 
Bouncing models can solve the horizon
problem with a long period of slow contraction ($a \sim (-t)^p$ with $0 < 
p < 1$) as opposed to inflationary models which undergo a period of 
quasi-exponential expansion. Models with a bounce play an important role 
in string cosmology (see \cite{PBB} and references therein).
It is therefore interesting to study how metric 
perturbations are generated in bouncing models paying particular 
attention to what happens close to the bounce, even in a simplified model.


In this paper we study a hydrodynamical toy model where the bounce is 
induced by a negative energy perfect fluid. Negative energy density fluids 
seem ideas quite wild - or desperate \cite{peebles}, but they have already 
been considered as 
toy models \cite{hwangnoh,PPN}. Moreover, negative energy 
fluid have been recently connected to brane models \cite{brane}, 
where the interaction 
between bulk and brane can induce a negative energy component in the brane 
through the projection of the $5$d Weyl tensor.

A two field description of a bounce is also very pedagogical for 
the treatment of cosmological perturbations through a bounce. In such a 
bounce, both the Hubble parameter $H$ and its time derivative $\dot H$ 
cross zero. Only the latter phenomena  
occurs in an expanding universe.  Indeed, the Hubble parameter changes slope during the 
transition from an accelerated to a decelerated expansion (from 
inflation to the standard big bang era, i. e. during reheating) and 
viceversa (from the matter dominated era to the present accellerated 
stage). The vanishing of $H$ is instead a peculiarity of a bounce. It 
turns out that the treatment of cosmological perturbations during a bounce 
is much more complicated than during a stage which only includes $\dot H = 
0$. 

The paper is organized as follows. In section 2 we introduce a novel class 
of bouncing universe. To our knowledge this class generalizes the case of 
a radiation dominated universe where the bounce is obtained with  
a massless scalar field with negative energy 
density \cite{peebles,hwangnoh,PPN}. In section 3 we study 
both in analytic and numerical ways
the evolution of gravitational waves in such a background. The equation 
can be solved analytically with the use of spheroidal functions, as Sahni 
did for gravitational waves in a radiation-matter transition 
\cite{sahni}. In section 4 we describe the problems in treating 
cosmological scalar perturbations during a bounce. In section 5 we study 
the evolution of scalar fluctuations and in section 6 we conclude.

\section{Background}

We consider a bouncing universe which satisfies the Einstein equation:
\be
H^2 = \frac{1}{3 M_{\rm pl}^2} \left[ \frac{\rho_+}{a^m} -
\frac{\rho_-}{a^n} \right] \,,
\label{hubble}
\ee
where $8 \pi G = M_{\rm pl}^{-2}$, $m < n$ and $\rho_+$, $\rho_-$ are 
constants. From this last relation it 
is obvious that $\rho_-$ is important only close to the bounce. 
The two fluids have ratios between pressure and energy densities $w_+ = 
m/3 -1$ and $w_- = n/3 -1$, respectively.

By assuming 
\be
n = 2 ( m - 1 ) \rightarrow  w_- = 2 w_+ + \frac{1}{3} 
\label{link}
\ee
the scale factor $a$ in terms of the conformal time is: 
\be
a (\eta) = \epsilon \left( 1 + \frac{\eta^2}{\eta_0^2} \right)^\alpha
\label{sfevolution}
\ee
with
\be
\epsilon = \left( \frac{\rho_-}{\rho_+} \right)^\frac{1}{n-m}
\ee

\be
\alpha = \frac{1}{1 + 3 w_+} = \frac{1}{n - m} = \frac{1}{m - 2}
\label{alpha}
\ee

\be
\eta_0^2 = 12 \alpha^2 M_{\rm pl}^2 \frac{\rho_-}{\rho_+^2} = 
\frac{12}{(n - m)^2} \frac{M_{\rm pl}^2}{\rho_+} \, \epsilon^{n-m}
\ee

This class of solutions covers many different cases of physical 
contraction, each of them bounced in an expansion phase by a negative 
energy density fluid with a different equation of state. The parameter 
$\epsilon$ regulates the magnitude of the scale factor at the bounce and 
the time scale $\eta_0$ sets the duration of the bounce.
Contraction driven by dust, radiation, free scalar field and ultra-stiff 
equation of state ($w_+ >> 1$) are given by $\alpha = 1, 1/2, 1/4$ and 
$\alpha \sim 0$, respectively. Since $w_- > w_+$ the bounce is 
unavoidable in Eq. (\ref{hubble}) and it does not depend on initial 
conditions.

Important relations are:
\be
w_+ = \frac{1}{3 \alpha} - \frac{1}{3} \,, w_- =  
\frac{2}{3 \alpha} - \frac{1}{3}
\ee

\be
{\cal H} = \frac{2 \alpha \eta}{\eta^2 + \eta_0^2} = \frac{2 \alpha 
x}{\eta_0 (1+x^2)}
\ee

\be
{\cal H}' = 2 \alpha \frac{\eta_0^2 - \eta^2}{(\eta^2 + \eta_0^2)^2} =
2 \alpha \frac{1 - x^2}{\eta_0^2 (1 + x^2)^2}
\ee

\be
\frac{a''}{a} = {\cal H}' + {\cal H}^2 = 
a^2 \frac{R}{6} = 
\frac{2 \alpha}{\eta_0^2 (1+ x^2)^2} \left[ 1 + (2\alpha - 1) x^2 \right]
\ee

\begin{figure}
\vspace{5.5cm}
\includegraphics{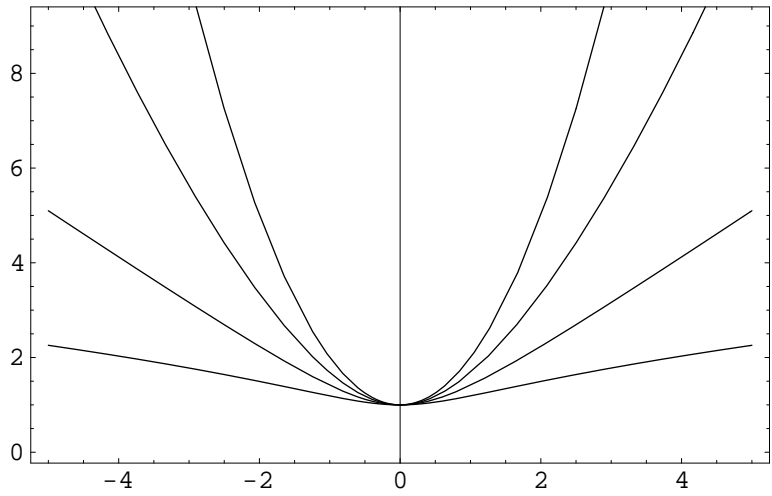}
\caption{The scale factor $a$ as function of $x$ for $\alpha = 1/4, 1/2, 
3/4, 1$ (from bottom to up, $\epsilon=1$). }
\label{qui}
\end{figure}

\begin{figure}
\vspace{5.5cm}
\includegraphics{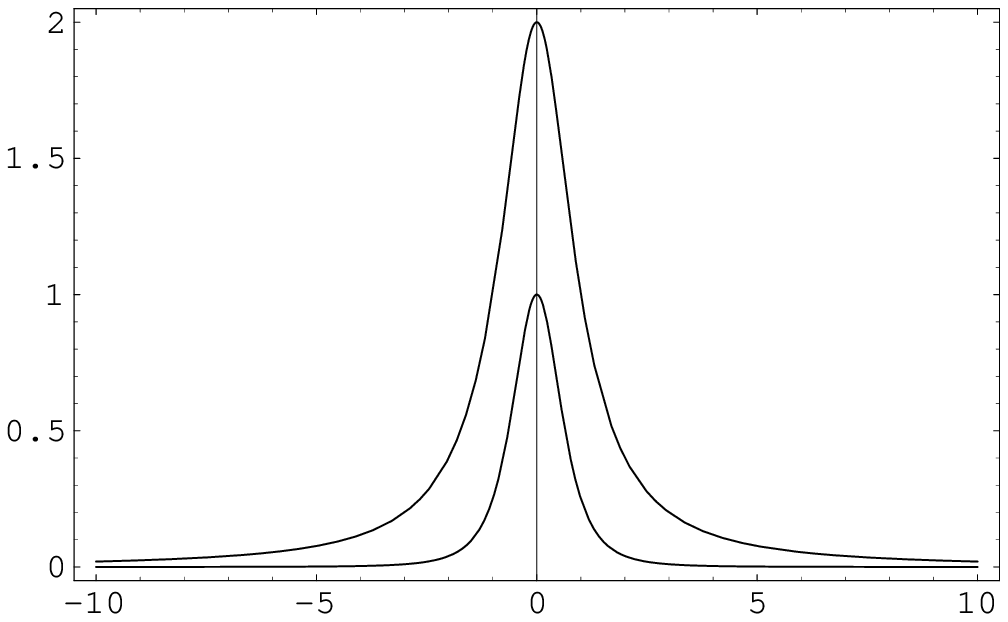}
\caption{The curvature term $a''/a$ as a function of $x$ for $\alpha=1/2 
\,, 1$ and $\eta_0=1$.}
\label{pippo}
\end{figure}

\begin{figure}
\vspace{5.5cm}
\includegraphics{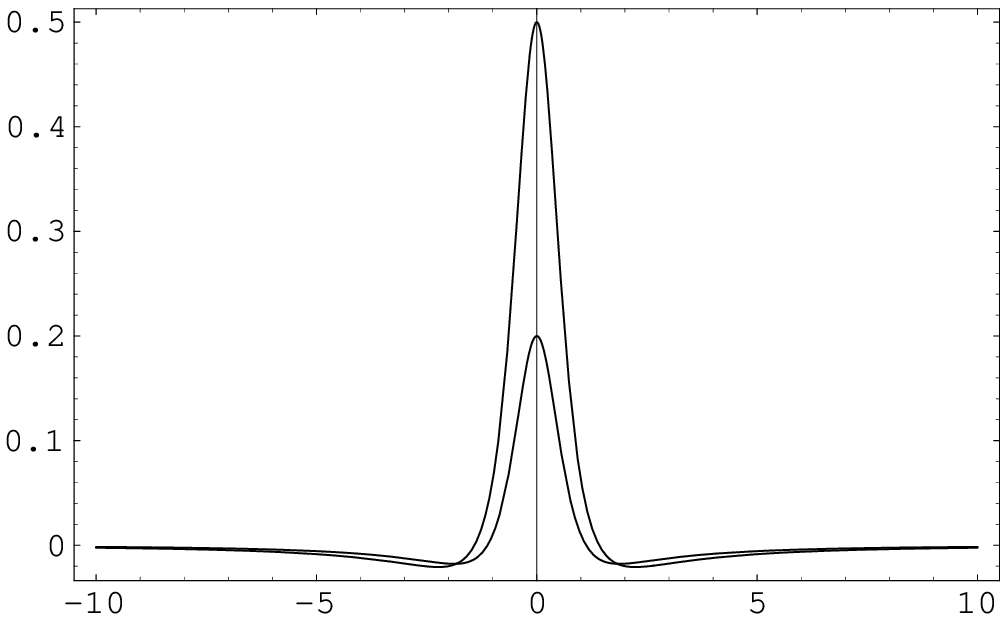}
\caption{The curvature term $a''/a$ as a function of $x$ for $\alpha=0.1 
\,, 0.2$ and
$\eta_0=1$.}
\label{topo}
\end{figure}

\begin{figure}
\vspace{5.5cm}
\includegraphics{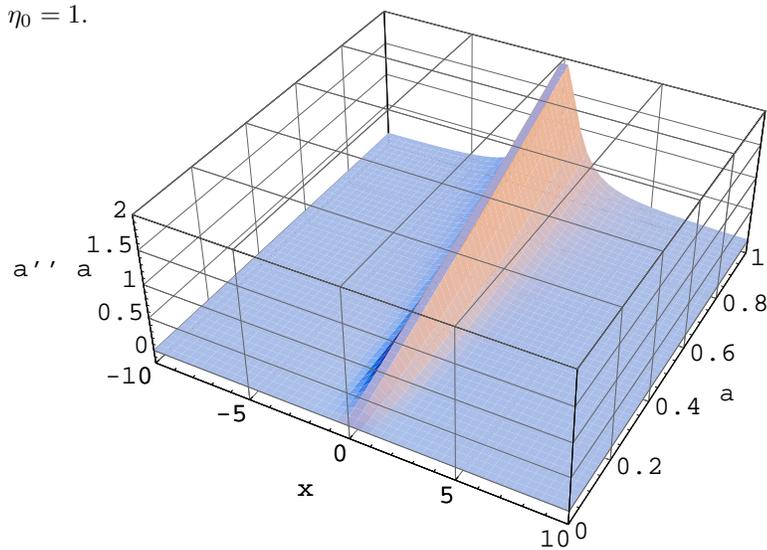}
\caption{The curvature term $a''/a$ as a function of $\alpha$ and $x$.}
\label{pluto}
\end{figure}
where $x \equiv \eta / \eta_0$.
We observe that $\rho_-$ cannot be simply related to the curvature of the 
spatial sections since $m \ge 3$ 
to avoid negative instability for fluid fluctuations \footnote{Scalar 
metric fluctuations have been studied during a bounce in \cite{GT} for 
a universe filled by a massive scalar field with closed spatial 
sections.}.
However, the above solution can be extendend to non flat 
spatial sections, as for the particular case $\alpha = 1/2$ \cite{brane}. 
The Hubble law in Eq. (\ref{hubble}) in general is:
\be
H^2 = \frac{1}{3 M_{\rm pl}^2} \left[ \frac{\rho_+}{a^m} -
\frac{\rho_-}{a^n} \right] - \frac{K}{a^2} \,,
\label{hubble2}
\ee
where $K = 0, 1, -1$ corresponds to flat, closed and open spatial 
sections. For $K=1$ the solution for the scale factor is:
\be
a (\eta) = 
\left( \frac{\rho_+}{6 M_{\rm pl}^2} \right)^\alpha \left[ 1 - c_1 \cos 
\left( 
\frac{\eta}{\alpha} \right) \right]^\alpha
\ee
where $c_1 = \sqrt{1 - 12 M_{\rm pl}^2 \rho_-/\rho_+^2}$.
For $K=-1$ the solution for the scale factor is:
\be
a (\eta) =
\left( \frac{\rho_+}{6 M_{\rm pl}^2} \right)^\alpha \left[ c_2 \cos{\rm 
h} \left(
\frac{\eta}{\alpha} \right) - 1 \right]^\alpha
\ee 
where $c_2 = \sqrt{1 + 12 M_{\rm pl}^2 \rho_-/\rho_+^2}$. Also for the 
open and closed solutions $\alpha$ is given by Eq. (\ref{alpha}).

\section{Gravitational Waves}

As usual, the amplitude of gravitational waves $h$ satisfies the following 
equation:
\be
h'' + 2 {\cal H} h' + k^2 h = 0 \,.
\label{gw}
\ee
By neglecting $k^2$ the infinite wavelength solution is:
\be
h \sim A + B \int \frac{d \eta}{a^2} = A + \frac{\eta_0}{\epsilon^2} \, B 
\,  
x \, _2 F_1 \left( \frac{1}{2}, 2 \alpha; \frac{3}{2}; - x^2 \right)
\ee

In particular, for a dust dominated collapse ($\alpha = 1$) we get
\be
h_{\rm mat} \sim A_{\rm mat} + B_{\rm mat} \, \frac{\eta_0}{\epsilon^2} 
\left( \frac{x}{2 (x^2 + 1)} + \frac{\arctan x}{2} \right) \,,
\ee
and for a radiation dominated collapse ($\alpha = 1/2$)
\be
h_{\rm rad} \sim A_{\rm rad} + B_{\rm rad} \, \frac{\eta_0}{\epsilon^2} \, 
\arctan x \,,
\ee
Instead, for a contraction driven by a free scalar field, we get:
\be
h_\phi \sim A_\phi + B_\phi \, \frac{\eta_0}{\epsilon^2} \, \arcsin 
{\rm h} \, x \,.
\ee

The presence of $\epsilon^2$ at the denominator indicates that the 
amplitude of gravitational waves blows up in the limit of singular 
bounces, i. e. $\epsilon \rightarrow 0$. 
This fact is not surprising and is related to the 
normalization condition between the two independent solutions for the 
modes $W (h, h^*) = i/a^2$. Fluctuations become singular in a singularity.

It is interesting to note that the B mode (odd with respect to $x 
\rightarrow - x$) tends to a constant far from the 
bounce, only for $\alpha \ge 1/2$. For $\alpha < 1/2$ the amplitude of 
gravitational waves grows outside the Hubble radius, even in the 
expanding phase. 
This result is not completely surprising, 
since for slow expansion ($a(\eta) \propto (\eta)^p$ with $p < 1/2$), 
free scalar fields (and henceforth gravitational waves) are a 
superposition of a constant and growing mode (but proportional to $k 
\eta$). Here one has this effects for a different range of parameters, 
probably because of the bounce.

These results indicate that it is potentially important to know the 
$k$ dependence of B, in 
order to predict the spectrum of gravitational waves far from the bounce. 
The contribution of the B mode to $h$ right at the bounce is zero in this 
infinite wavelength approximation.

\begin{figure}
\vspace{4.5cm}
\includegraphics{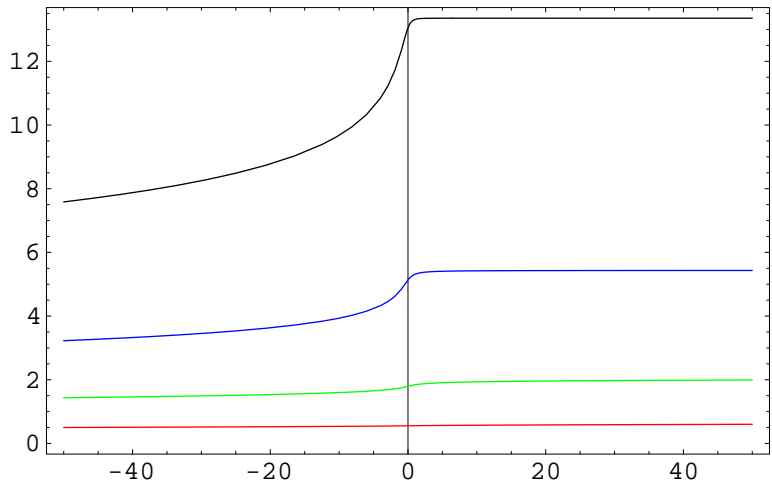}
\caption{Behaviour of a long-wavelength gravitational wave across the
bounce. The plots are for $\alpha = .1, .25, .5. 1$ (from
bottom to top).}
\label{gw_figure}
\end{figure}

With the evolution of the scale factor in Eq. (\ref{sfevolution}), the 
equation for $h$ can be solved {\em analytically}. The trick is to put the 
above equation in 
the form of the {\em differential equation of spheroidal wave function} 
\cite{bateman}. 
It is useful to introduce then:
\be
x = \eta/\eta_0  \quad  \tilde h = h (1 + x^2)^{\alpha - 1/2} \,.
\ee

After a little algebra one gets:
\be
\frac{d^2 \tilde h}{d x^2} + \frac{2 x}{1 + x^2} \frac{d \tilde h}{d x}  
+ \left[ k^2 \eta_0^2 - \frac{2 \alpha (2 \alpha - 1) }{1 + x^2} 
+ \frac{(2 \alpha - 1)^2}{(1 + x^2)^2} \right] \tilde h = 0 \,.
\label{tilde}
\ee
We note that the above equation for $k \eta_0 = 0$ reduces to the Legendre 
differential equation \cite{bateman,abramowitz}.
The solution to Eq. (\ref{tilde}) is given by the radial oblate spheroidal 
function \cite{abramowitz}. The above equation can be obtained by 
the transformation $z \rightarrow \pm i x$ in the equation in prolate 
coordinates \cite{bateman,abramowitz,sahni}:
\be
(1 - z^2) \frac{d^2 \tilde h}{d z^2} - 2 z \frac{d \tilde h}{d z}
+ \left[ \gamma^2
(1 - z^2) + \nu(\nu + 1)
- \frac{\mu^2}{(1 - z^2)} \right] \tilde h = 0 \,,
\ee
where $\mu \,, \nu \,, \theta$ are numerical coefficients. In many books 
only the case with integer $\mu \,, \nu$ is treated 
\cite{bateman,abramowitz}, however $\mu \,, \nu 
\,, \gamma$ can be even complex numbers \cite{falloon}.

The solution for $h$ is:
\be
h = \epsilon \sqrt{\frac{\eta_0 \gamma}{2}} (1 + x^2)^{1/2 - \alpha} 
S^{2 \alpha - 1 \, (3)}_{2 \alpha - 1} (x, \gamma) \,.
\label{solution}
\ee
where (the following series is convergent only when $|z| > 1$ 
\cite{falloon}, to have convergence at all times one needs a different 
expression):

\be
S^{\mu \, (3)}_{\nu} (x, \gamma) = \left( \frac{\pi}
{2 \gamma x}
\right)^{1/2} \left( \frac{x^2}{x^2 + 1} \right)^{\mu/2} 
\sum_{r=0}^{\infty} \frac{a^{\mu}_{\nu \,, 
r}(\gamma)}{s^{\mu}_{\nu}(\gamma)} 
H^{(1)}_{\nu+1/2+2 r} (\gamma z)
\ee

\be
\gamma = - i k \eta_0 \,, \quad z = - i x \,, \quad 
\gamma z = - k \eta
\ee

\be
\mu = \nu = 2 \alpha - 1
\ee

\begin{figure}
\vspace{4.5cm}
\includegraphics{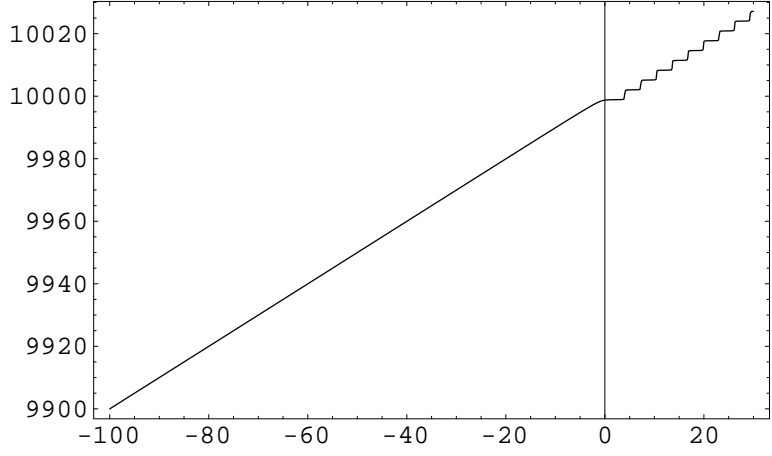}
\caption{Behaviour of a the phase of a long-wavelength gravitational wave
across the bounce for $\alpha = 1$.}
\label{gw_phase}
\end{figure}

The normalization factor is fixed by requiring that for 
$x \rightarrow \infty$ (and then $|k \eta| \rightarrow \infty$):
\be
h \sim \frac{e^{-i k \eta}}{a \sqrt{2 k}} \,.
\ee   

\be
\lim_{x \rightarrow \infty} S^{\mu \, (3)}_{\nu} (x, \gamma) =
\left( \frac{\pi}{2 \gamma x} \right)^{1/2} H^{(1)}_{\nu+1/2} (\gamma z)
\label{largex}
\ee

In order to understand the behaviour for large wavelengths 
we follow Sahni 
\cite{sahni} and we take the limit for $\gamma \rightarrow 0$ 
\footnote{This limit identifies the infrared modes at the bounce, i. e. 
$k^2 << a''/a$ at $\eta \sim 0$.} \cite{bateman}:
\be
\lim_{\gamma \rightarrow 0} S^{\mu \, (3)}_{\nu} (x, \gamma) =
\gamma^{\nu/2} \left( \#_1 P^\mu_\nu (x) + \#_2 Q^\mu_\nu (x) \right) + 
\frac{\#_3}{\gamma^{\nu+1}}  Q^\mu_\nu (x)  
\ee
where $P^\mu_\nu (x) \,, Q^\mu_\nu (x)$ are respectively the associate 
Legendre functions of the first and second kind \cite{bateman} 
and $\#_1,\#_2,\#_3$ are numerical coefficients.
Gravitational waves have finite amplitude right at the bounce since 
$P^\mu_\nu (0)$, $Q^\mu_\nu (0)$ are finite (see Eqs. 
8.6.1 and 8.6.2 in \cite{abramowitz}).
This leads to a spectrum for gravitational waves in the infrared:
\be
\lim_{\theta \rightarrow 0} |h|^2 \sim 
\frac{1}{k^{2 \nu +1}} \,.
\label{smalltheta}
\ee
The above result is in formal agreement with Sahni \cite{sahni}. 

We note that the 
phases of gravitational waves are not constant while outside the Hubble 
radius, as shown in Fig. (\ref{gw_phase}). Phases are locked to a constant 
value when the wavelength of a mode exceeds the Hubble radius during 
inflation: this 
phenomenon has been dubbed "decoherence without decoherence" \cite{PS}.
Models based on a contraction which aim to produce a scale invariant 
spectrum seem to not have this peculiarity. 

\begin{figure}
\vspace{2.5cm}
\includegraphics{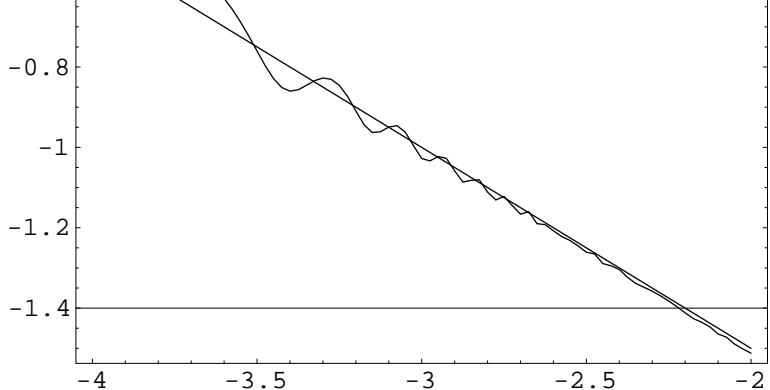}
\caption{A snapshot of the gravitational wave amplitude as a function 
of $log_{10} k \eta_0$ at $x=30$ for $\alpha=1/2$. The other 
detail of the 
simulation are $x_i = - 10^5$, $\epsilon = 10^3$ and a initial vacuum 
spectrum for gravitational waves. The linear fit is $-0.8 - 0.5 
(log_{10} k \eta_0 + 3.4)$.} 
\label{spectrumgwrad} 
\end{figure}

\begin{figure}
\vspace{4.5cm}
\includegraphics{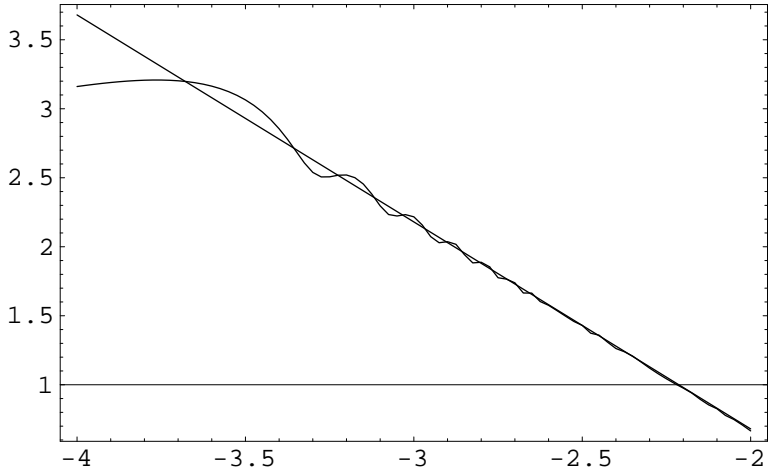}
\caption{A snapshot of the gravitational wave amplitude as a function
of $log_{10} k \eta_0$ at $x=30$ for $\alpha=1$. The other details the 
same of the previous figure. The linear fit is $2.78 - 
1.5 (log_{10} k \eta_0 + 3.4)$.}
\label{spectrumgwdust}
\end{figure}

The final spectrum of gravitational waves far from the bounce is scale 
invariant for a dust contraction ($\alpha = 1$), and it is a vacuum 
spectrum for a radiation contraction ($\alpha = 1/2$), as we can see from 
Figs. (\ref{spectrumgwrad},\ref{spectrumgwdust}). 

\vspace{1cm}

\section{Equations for Scalar Perturbations}

We now review the equations of motion of scalar perturbations in the
longitudinal gauge \cite{MFB} with two hydrodynamical fluids. The 
energy and momentum
constraints are respectively:
\be
- \Delta \Phi + 3 {\cal H} \left( \Phi' + {\cal H} \Phi \right) =
4 \pi G a^2 \left( \delta \rho_+ - \delta \rho_- \right)
\label{enconstr}
\ee
\be
\Phi' + {\cal H} \Phi = 4 \pi G \frac{a^2}{k} \left[ \rho_+ (1 + w_+) v_+
- \rho_- (1 + w_-) v_- \right]
\ee
The spatial part of the Einstein equations lead to:
\be
\Phi '' + 3 {\cal H} \Phi' + \left[ 2 {\cal H}'
+ {\cal H}^2 \right] \Phi = 4 \pi G a^2 \left( \delta p_+ - \delta p_- \right)
\,,
\label{eqphi}
\ee

Usually Eq. (\ref{enconstr}) times the total sound speed $c_s^2$: 
\be
c_s^2 \equiv \frac{\dot p}{\dot \rho} = \frac{\dot p_+ - \dot p_-}{\dot 
\rho_+ - \dot \rho_-}
\label{soundspeed}
\ee
is subtracted by Eq. (\ref{eqphi}) to obtain:
\be
\Phi '' + 3 (1+c_s^2) {\cal H} \Phi' + \left[ c_s^2 k^2 + 2 {\cal H}'
+ {\cal H}^2 + 3 c_s^2 {\cal H}^2 \right] \Phi = 4 \pi G a^2 \delta S
\,,
\label{eqphitot}
\ee
where $\delta S$ are the total non adiabatic pressure perturbations:
\be
\delta S = \sum_i \delta p_i - c_s^2 \sum_i \delta \rho_i \,.
\ee

In problems with a bounce the sound speed $c_s^2$ as defined in Eq. 
(\ref{soundspeed}) becomes singular \footnote{Recently a paper which 
studies numerically the spectrum of
$\Phi$ fluctuations during a fictious bounce driven by a single
perfect fluid appeared \cite{CDC}. We think that the
toy model used in \cite{CDC} is inconsistent.
Infact the equation of motion of $\Phi$ used in \cite{CDC} contains
a regular and {\em ad hoc} quantity which substitues the physical
$c_s^2$, which becomes singular during the bounce.}.
This is connected with the violation of the null energy 
condition (henceforth NEC) $\rho + p \ge 0$, which occurs not right at the 
bounce ($\eta = 0$), but when 
\be
\dot H = - \frac{1}{2 M_{\rm pl}^2} (\rho + p) = - \frac{1}{2 M_{\rm pl}^2}
\sum_i (1+w_i) \rho_i
\ee 
vanishes, i.e. when $\eta = \eta_0/2$ in this class of models.

This problem has been already encountered in the study of cosmological 
perturbations during reheating \cite{reheating1}. In a scalar field 
driven cosmology $\dot H$ vanishes when the field bounces, i.e. 
$\dot \phi =0$, otherwise it remains always negative. The equation for 
$\Phi$ has singularities \cite{reheating1}.
In \cite{reheating1,reheating2} the problem was approached by using the 
Mukhanov variable 
$Q ( = \delta \phi + \dot \phi \, \Phi /H)$ whose evolution is regular 
during the oscillation of a scalar field, despite the NEC violation. The 
regularity of the equations which involve the $Q$s persist also in the 
multifield case \cite{reheating2}.

When a bounce is present the equations governing the evolutions of the 
$Q$s become singular in $H=0$ (see \cite{reheating2} for scalar fields). 
This would be true also for perfect fluids. Infact, for a system 
of perfect fluids characterized by $w_F$ and $c_F^2$, the Mukhanov 
variables $Q_F$s would satisfy:
\begin{eqnarray}
\ddot Q_F + 3 H \dot Q_F &+& \left[ c_F^2 \frac{k^2}{a^2} + \frac{3}{2} 
\dot H (1 - w_F) + \frac{9}{4} H^2 (1 - w_F^2) \right] Q_F = \nonumber \\ 
&& \frac{1}{H} \times ( Qs, \dot Qs \quad {\rm of} \quad {\rm other} \quad 
{\rm components})
\,.
\end{eqnarray}
We note that in presence of just one component,
only a dust contraction can generate a scale invariant spectrum, as for 
the scalar field case \cite{si1}.

We therefore turn to different variables in the next section.

\section{Evolution of Scalar Perturbations}

In the appendix of \cite{PPN} Peter and Pinto-Neto suggest a possible way 
to deal with metric perturbations during a bounce. They suggest to split 
the Newtonian potential $\Phi$ in two components
\be
\Phi = \Phi_+ + \Phi_-
\ee
where each one satisfies:
\be
\Phi_i'' + 3 (1+w_i) {\cal H} \Phi_i' + \left[ w_i k^2 + 2 {\cal H}' 
+ {\cal H}^2 + 3 w_i {\cal H}^2 \right] \Phi_i = 0 
\label{eqphii}
\ee
with $i=+\,,-$. This splitting is valid under the assumption that non 
adiabatic pressure perturbations in the two components are absent:
\be
\delta p_i = c_i^2 \delta \rho_i  \quad {\rm with} \quad c_i^2 = w_i
\label{assump}
\ee
It is important to note that the above equation(s) are 
{\em regular} and {\em uncoupled}. 

It is very interesting to see that the contribution of $\Phi_-$ to $\Phi$ 
decays in the expanding phase. The reason for this behaviour is that 
$\Phi_-$ does not have a constant mode in the long wavelength limit 
far from the bounce, as instead $\Phi_+$ does. In order to see this, it is 
sufficient to evaluate
\be
2 {\cal H}' + {\cal H}^2 + 3 w_i {\cal H}^2 = - 3 (w - w_i) {\cal H}^2 \,.
\ee
For $i = -$ we get: 
\be
a^2 \frac{\rho_+}{3 M_{\rm pl}^2} (n - m) \,,
\ee
which is a strictly positive quantity at all times.
For $i = +$ we get a vanishing quantity at leading order far
from the bounce:
\be
w_+ \left( 3 {\cal H}^2 - a^2 \frac{\rho_+}{M_{\rm pl}^2} \right) + w_- 
a^2 \frac{\rho_-}{M_{\rm pl}^2} \,.
\ee

\begin{figure}
\vspace{4.5cm}
\includegraphics{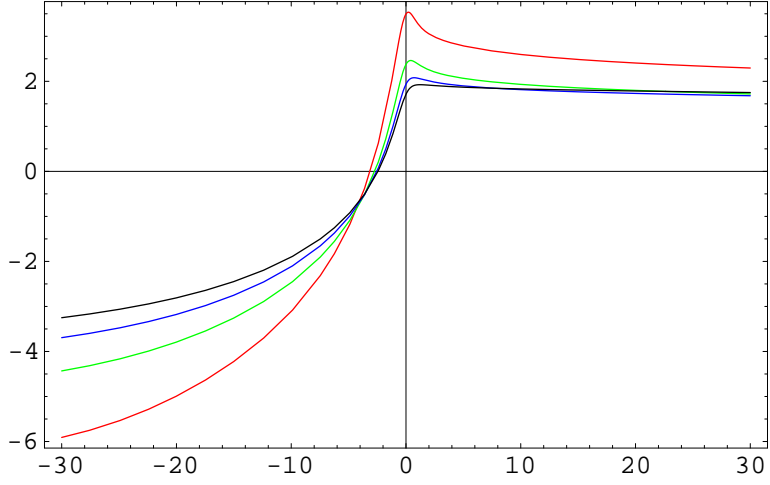}
\caption{Behaviour of a long-wavelength mode of $\Phi_-$ across the
bounce. After a transient decay around the bounce, $\Phi_-$ does 
not stabilize at a constant value, as also
seen analytically. The plots are for $\alpha = 0.1, 0.25, 0.5, 1$ (from
bottom to top after the bounce).}
\label{decaying}
\end{figure}

The above result implies that the growing mode of $\Phi_-$ during the 
contracting phase does not match to the growing mode of 
$\Phi$ in the expanding phase. This result is confirmed by numerical 
analysis, as shown in Fig. (\ref{decaying}).

Therefore, the $-$ component is 
important only close to the bounce both at the background level and at the 
linear level. 

We now turn to the analysys of $\Phi_+$ and we repeat the previous 
treatment for long wavelengths. For $i = +$ we get a vanishing 
frequency at leading order far from the bounce:
\be
w_+ \left( 3 {\cal H}^2 - a^2 \frac{\rho_+}{M_{\rm pl}^2} \right) + w_-
a^2 \frac{\rho_-}{M_{\rm pl}^2} \,.
\ee
Now we would like to recast Eq. 
(\ref{eqphii}) in a spheroidal equation form. By introducing: 
\be
u_i = (1 + x^2 )^{-\delta} \Phi_i
\ee
where $\delta = 1/2 - 3 \alpha (1 + w_i)/2$. The equation for $u_i$ is:
\be
u_i '' + \frac{2 x}{1 + x^2} u_i ' + \left[ w_i \, k^2 \eta_0^2 + 
\frac{b_i}{1 + x^2} + \frac{c_i}{(1 + x^2)^2} \right] u_i = 0
\,.
\ee
where
\be
b_i = 4 \alpha ( \alpha (1+3 w_i) - 1) + 6 \alpha \delta (1 + w_i)
\ee
\be
c_i = 4 \alpha ( 1 + 2 \delta ) + 4 \delta^2
\ee
When $i=+$, we get $b_+ = - 2 \alpha (2 \alpha + 1)$, $c_+ = 4 \alpha 
(1+\alpha)$, leading to:
\be
\nu_+ = 2 \alpha \quad \mu_+ = \sqrt{4 \alpha (1 + \alpha)} \,.
\ee
The solution is therefore
\be
\Phi_+ = \frac{\cal N}{(1 + x^2 )^\alpha} S^{\mu_+ \, (3)}_{2\alpha} (x, 
\gamma) 
\ee
where ${\cal N} \sim {\cal O} (\gamma^{-1/2})$ is fixed by imposing:
\be
\Phi \sim \frac{1}{k^{3/2}}
\ee
on short wavelengths far from the bounce. Indeed, 
\begin{eqnarray}
\lim_{- k \eta \rightarrow \infty} 
\lim_{x \rightarrow \infty} \Phi_+ &=& \lim_{- k \eta \rightarrow 
\infty} \frac{\cal N}{(1 + x^2 )^\alpha} 
\left( \frac{\pi}
{2 \gamma x}
\right)^{1/2} H^{(1)}_{2 \alpha + 1/2} 
(\gamma x)
\nonumber \\
&=& \frac{\cal N}{(1 + x^2 )^\alpha} \left( 
\frac{1}{\gamma^2 x^2}
\right)^{1/2} e^{-i k \eta}
\end{eqnarray}

\begin{figure}
\vspace{4.5cm}
\includegraphics{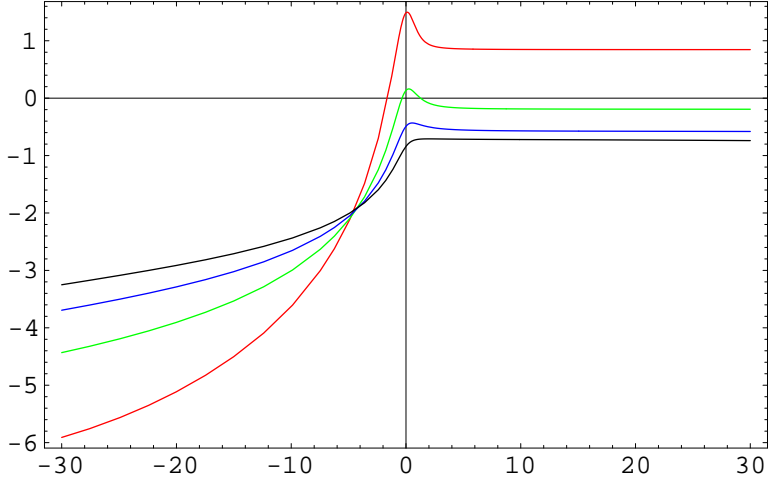}
\caption{Behaviour of a long-wavelength mode of $\Phi_+$ across the 
bounce. After a transient decay around the bounce, 
$\Phi_+$ stabilizes at a constant value, as also 
seen analytically. The plots are for $\alpha = 0.1, 0.25, 0.5., 1$ (from 
bottom to top after the bounce).}
\label{growing}
\end{figure}

For large wavelengths, by using Eq. (\ref{smalltheta}) one has:
\be 
\lim_{\gamma \rightarrow 0} |\Phi_+|^2 \sim
\frac{1}{k^{3+4\alpha}} \,.
\ee
A nearly scale invariant spectrum for $\Phi_+$ emerges in the expanding 
phase for $\alpha \sim 0$. 
For all the other relevant values of $\alpha$ the spectrum 
for $\Phi_+$ in the expanding phase seems too red to agree with 
observations (for $\alpha = 1/2$ $\Phi_+ \sim k^{-5/2}$ in agreement 
with \cite{PPN}).


%


\section{Discussions and Conclusions} 

We have studied cosmological perturbations through a non-singular 
bounce in a simple class of with two perfect fluid toy models. 
A normal fluid drives the 
contraction/expansion, while a negative energy density fluid models the bounce.
The equations of state of the two fluids are linked in a particular way 
described by Eq. (\ref{link}).

The study of scalar metric fluctuations during a non-singular bounce is 
rather tricky, but it can be approached, even analytically, as this 
example shows.
Following \cite{PPN}, one could find variables whose evolution 
equation is regular through the bounce. We do not think that the approach 
proposed in \cite{PPN} is
unique in order to evolve cosmological scalar perturbations through a
bounce. One could define a {\em fictious} total sound speed ${\tilde
c_s^2}$ which is regular through the bounce and asympotically behaves like
the sound speed of the dominant component. There is a
possibility that in this way also the equation of motion for
isocurvature perturbations is regularized. It remains to study the
coupling of $\Phi$ to isocurvature perturbations case by case.
We observe
that this trick of a {\em fictious} total sound speed has been used in
\cite{CDC} in order to evolve $\Phi$ and ignoring isocurvature
perturbations. This of course is equivalent to study only half of the
system and neglect the possibile interaction between isocurvature and 
adiabatic mode.

The metric fluctuations induced by the fluid which drives the contraction 
$\Phi_+$ go through the bounce and seed the constant (growing) mode in the 
expansion. The metric fluctuations induced by the exotic fluid $\Phi_-$ 
are important only close to the bounce and decay with time in the expanding 
phase. However, we note that the latter have a spectrum which is steeper 
in the infrared than the constant mode. This means that the largest piece 
of $\Phi$ in the contracting phase matches to a decaying mode in the 
expanding phase \cite{BF}. The picture which emerges seems the 
following: in presence of a second entity which is responsible for the 
bounce, both the growing and decaying mode of metric fluctuations 
seeded by the dominant fluid which drives the 
contraction match to the growing mode of $\Phi$ in the expanding phase. 

In the expanding phase, after the bounce, gravitational waves have a scale 
invariant spectrum for a dust contraction as originally discovered by 
Starobinsky \cite{staro}. This is also what we find here (see also 
\cite{si1,si2}).
In this two field model, after the 
bounce, scalar fluctuations with a (nearly, slightly red-tilted) scale 
invariant spectrum are generated by a very slow contraction, 
corresponding to a contracting fluid with an ultra-stiff equation of state 
($w_+ >> 1$). In this picture, the scale invariant spectrum for sub-Hubble 
metric fluctuations $\Phi$ is adiabatically transferred to 
super-Hubble wavelengths by the slow contraction, but remains imprinted 
in the subsequent expanding phase because of the bounce induced by a 
second entity. The amplitude of $\Phi_+$ in agreement with observations 
could be obtained by tuning the parameters of the model. In the 
same way $\Phi_-$, hopefully, should be kept in the linear regime during 
the bounce. This possibility could be of interest 
for the Ekpyrotic scenario \cite{Ekpyrotic}, without the need of a 
singularity. The model obtained would be similar to the idea of hybrid 
inflation in the context of inflationary models. As in hybrid inflation 
\cite{hybrid} one field drives inflation and the other terminates the 
accelerating stage, in this case one component would drive the contraction 
and the other would be responsible for the bounce.

It is possible that the above conclusions are due to the particular 
relation among the equations of state of the two fluids or to the time 
simmetry of the model, but this seems unlikely. 
The absence of intrinsic non adiabatic pressure perturbations 
(i. e. the assumption of two perfect fluids) may be a more important 
issue. The danger could be a coupling (it may be just of gravitational 
origin) of $\Phi_-$ to $\Phi_+$. In such a case a scale invariance of 
metric fluctuations in the expanding phase may be very difficult to 
obtain. In any case this toy model shows how the physics at the bounce can 
alter the predictions about cosmological perturbations also in the PBB 
scenario \cite{PBB}.

It remains to connect this study with a prescription for matching 
perturbation through a non-singular bounce in a multifield case.


\vspace{1cm}

{\bf Note Added}: While this paper was almost completed, two papers which 
studied metric perturbations during a bounce appeared \cite{GGV,TTS}. 
Gasperini, Giovannini and Veneziano \cite{GGV} studied a single field 
non singular bounce induced by a non local potential and found that the 
growing mode of $\Phi$ in the contracting phase 
(not supported by a growing mode of $\zeta$) 
does not match to the growing of $\Phi$ in the expanding phase 
\cite{DM,BF}. Previous studies with a 
bounce induced by loop corrections (with or without a local potential) 
\cite{CCH,TBF} agree with \cite{GGV}. 
This result could be related to the single field assumption. 
Tolley, Turok and Steinhardt \cite{TTS} studied cosmological 
perturbations through a singular bounce.

\vspace{1cm}

{\bf Acknowledgments}

I would like to thank Robert Brandenberger for many comments on 
earlier version of this draft. I am grateful to Gianpaolo Vacca 
for his kind help with numerics. I would like to thank Patrick Peter 
and Peter Falloon for clarifications and discussions.

\end{document}